\documentclass[floatfix,10pt,onecolumn,showpacs,amsmath,amssymb]{revtex4}
\usepackage{epsf}
\usepackage{graphicx}  
\usepackage{dcolumn}   
\usepackage{bm}        
\usepackage{color}

\newcommand{\be}{\begin{equation}}
\newcommand{\en}{\end{equation}}
 \newcommand{\bea}{\begin{eqnarray}}
 \newcommand{\ena}{\end{eqnarray}}
  \newcommand{\sch}{Schwarzschild}

\begin{document}

\title{Tail wavelets in the merger of binary compact objects}
\author{ Kai Lin$^{1,2} $, Wei-Liang Qian $^{2,3,4}$, Xilong Fan$^{5}$, and Hongsheng Zhang$^{6,7}$\footnote{Electronic address: sps\_zhanghs@ujn.edu.cn}}
\affiliation{
$^{1}$ Institute of Geophysics and Geoinformatics, China University of Geosciences, 430074, Wuhan, Hubei, China\\
$^{2}$ Escola de Engenharia de Lorena, Universidade de S\~ao Paulo, 12602-810, Lorena, SP, Brazil\\
$^{3}$ Faculdade de Engenharia de Guaratinguet\'a, Universidade Estadual Paulista, 12516-410, Guaratinguet\'a, SP, Brazil\\
$^{4}$ School of Physical Science and Technology, Yangzhou University, 225002, Yangzhou, Jiangsu, China\\
$^{5}$ School of Physics and Technology, Wuhan University, 430072, Wuhan, Hubei, 430205, China\\
$^{6}$ School of Physics and Technology, University of Jinan, 336 West Road of Nan Xinzhuang, Jinan, Shandong, 250022, China\\
$^{7}$ State Key Laboratory of Theoretical Physics, Institute of Theoretical Physics, Chinese Academy of Sciences, Beijing, 100190, China
}

\date{ \today}

\begin{abstract}
 We present a model for tail wavelets, a phenomenon also known as ¡°echo¡± in the literature. The
tail wavelet may appear in signal reconnaissances in the merger of binary compact objects, including
black holes and neutron stars. We show that the dark matter surrounding the compact objects lead
to the speculated tail wavelet following the main gravitational wave (GW). We demonstrate that the radiation pressure
of the main wave is fully capable of pushing away the the surrounding matter to some altitude,
and splashing down of the matter excites the tail wavelet after ring down of the main wave. We
illustrate this idea in a simplified model, where numerical estimations are carried out concerning
the specific distribution of the dark matter outside the black hole horizon and the threshold values
in accordance with observations. We study the full back reaction of the surrounding dark matter to
the metric, and find that the effect is insignificant to the tail wavelets. We find the fine difference
between the tail wavelets of a dressed black hole and a bare one. We demonstrate that the tail
wavelet can be a natural phenomenon in frame of general relativity, without invoking any modified
gravities or quantum effects.
\end{abstract}

\pacs{04.20.-q, 04.70.-s}
\keywords{gravitational wave; dark matter; tail wavelet}

\preprint{arXiv: }
\maketitle

{\textit{I.~Introduction}}

The observations of gravitational waves (GWs) from binary compact objects symbolizes the beginning of an era of GW astronomy as a branch of observational astronomy with considerable precision.
Regarding fundamental physics, we finally have the experimental observables at our disposal for testing whether a quantum theory of gravity is plausible.
The GW is an incisive tool to study the structure of the black holes and other compact objects.
In particular, it may unwrap the enigma about the quantum structure of the spacetime beyond the horizon.
In regard to the quantum effects, it is believed that the spacetime is no longer a smooth manifold.
It may possess a specific non-trivial structure when the quantum fluctuations become significant \cite{MTW}.
In fact, the structure of the black hole horizon has been suggested to solve the black hole information problem, for example, in terms of the fuzzball model \cite{fuzzb} and the firewall model \cite{firewall1, firewall2}.
Such a speculated structure may be detected by the GWs emitted from the binary black holes \cite{GWs}.

In these models, the horizon is not a smooth sphere, but a shell with finite thickness.
 As a result, tail wavelets referred  as ``echoes" may appear after the main GWs have been emitted \cite{echo1, echo2, echo31, echo32}. The waveform of the echo model suggested in \cite{echo1} may be ruled out by analysis in \cite{noecho1, noecho2}. However, the tail wavelet may be in a different form compared with \cite{echo1}. One can deal with this problem from an opposite direction, that is, to find what will happen if there is completely no signal in the data after the main wave \cite{noecho1, noecho2}.  An detailed analysis in \cite{renjing} demonstrates that signals appear for several GW events with a p-value of order 1\% or sometimes significantly less. More approaches for signal reconnaissance of the tail wavelets are presented in \cite{duchen}.
In this context, it turns out to be somewhat a surprise when the tail wavelets are also observed in GW events of the binary neutron stars, besides those of binary black holes.
While explanations are given in terms of the resulting black hole remnant \cite{NStarwave}, the observed events also imply the possibility that the tail wavelet may be related to specific generic properties of the binary system
        rather than the near-horizon quantum structure \cite{NStarwave2}.
       In fact, it would be quite challenging to interpret the appearance of tail wavelets in terms of the properties of the horizon if the latter might not exist in the first place.

Following the above train of thought, we proceed by discussing some of the generic properties of the compact objects, which might potentially cause the tail wavelets.
Generally, a black hole/neutron star is surrounded by dark matters, whose density is much higher than that of the average value in the universe.
We will show later that the dark matter distributed in the vicinity of the compact object can provide a mechanism for the observed tail wavelets.
Further quantitative studies of the wavelets may be helpful to extract the distribution of the dark matter around the compact objects.
We note that in the literature there exist extensive studies on the distribution of dark matter around celestial bodies.
Some pioneer works \cite{piwork1, piwork2} have explored specific properties of the dark matter distributions in the halos of various mass scales.
A remarkable feature of the distribution of the dark matter is its universality.
Halos of different scales, namely, from the mass of the celestial body to that of a cluster of galaxies, all possess similar distribution \cite{nfw1, nfw2}.
For the supermassive black holes in elliptical galaxies and barred/ordinary spiral galaxies, such as our Milky Way, the density profile is studied quantitatively \cite{sbh1, sbh2}.
For the stellar mass black holes, direct observation concerning the dark matter halo is still absent. The distributions of dark matters around neutron stars, whose space environments are very similar to  stellar mass black holes, are studied in detail \cite{disneu1, disneu2, disneu3}. Regarding the universal distribution of the dark matter halos, it is reasonable that binary compact stars/black holes are immersed in dark matter halos.

Our proposed scenario is described as follows.
The passage of the main GWs from the binary system merger pushes up the dark matters to a certain altitude.
After the initial waves traverse through, the matter falls back towards the black hole region and subsequently produces the tail wavelets.
The relevant physical quantities are the energy density, pressure, energy flux of the GWs.
For linear plane GWs with frequency $\omega$ and amplitude $A$, the energy density $\rho$ is found to be \cite{MTW}, $\rho=\frac{\omega^2}{32\pi}(A_{+}^2+A_{\times}^2)$,
while the energy flux $b$ in the direction of propagation reads,$b=\frac{\omega^2}{32\pi}(A_{+}^2+A_{\times}^2).$
The above expressions are the same with the Landau-Lifshitz pseudo-tensor for GWs.

In a general case, the stress-energy tensor for the GWs can be defined as a second order perturbation of the Einstein tensor.
The general definition of GWs in the strong field region leads to some uncertainties, which depend on the specific choice of the background for the GWs in question.
For example, the obtained waves possess different forms when one expands the metric about a Minkowski spacetime as compared to a~\sch~one.
Nonetheless, the relation between power per area $S=b$ and the pressure $P$ satisfies
\be
P=S/c, \label{radi}
\en
for massless fluctuations (gravitons). Here we recover the luminous velocity $c$. It is noted that Eq.(\ref{radi}) is valid for GWs in a general sense, independent of the background metric.
As the total energy radiated from a binary object can be calculated, one thus obtains the corresponding pressure.
In what follows, we estimate the order of magnitude of the pressure by considering the GW150914 event as an example.
The system emits an amount of energy equivalent to 3 solar mass in about 0.1 seconds.
The GW emission is ``directional". However, its directivity is much weaker than electromagnetic radiation, since the lowest order radiation is of the quadrupole.
Thus, if one omits the GWs' directivity, one may approximate the emission as isotropic and estimate the pressure of GW at the horizon by,
\be
P=\frac{w}{4\pi r_h^2 c}\sim \frac{3\times 10^{49}}{4\pi (62\times 3\times 10^3)^2\times 3\times 10^8}=2.3\times 10^{29} {\rm Pa}.
\en
As for comparisons, the pressure at the center of a neutron star can reach $10^{33}~{\rm to}~ 10^{34}$Pa.
So the estimated value of the GW pressure is found to be comparable to those of the interior of stellar systems.

At the region near horizon, the Newton gravity fails. We just make an estimate to clear which force, the pressure or attraction, dominates for an object around the black hole. Assuming an object whose mass is 1kg and area is 1 $m^2$. For GW150914, it is attracted by the black hole with force at the horizon $F=G\frac{m_1m_2}{r^2}=2\times 10^{11}N$. At same time, the pressure it sensed is $F'=2.3\times 10^{29}N$ when the main wave passes. It is clear  the repulsive force is much lager than than attractive force for this object. Both the repulsive force and attractive force decrease with $r^{-2}$. Thus, at large distance, the ratio between   the repulsive force and attractive force is exactly $F'/F$, if the wave is completely absorbed by the object.

Furthermore, we assume that the surrounding matters can only absorb a tiny fraction of GW.
A rigorous treatment of the scattering between the graviton and matter demands a full-fledged theory of quantum gravity, which is still unknown.
Tree level calculations may also be entirely plagued by loop corrections.
Nevertheless, some preliminary studies on the scattering of gravitons have been carried out, see for example \cite{hol1, hol2}.
By very generic arguments, similar to those for earlier approaches of quantum physics, one can estimate the lower bound of total cross-section of a graviton scattered by a matter particle $\sigma\sim 10^{-68} {\rm m}^2$ \cite{dyson}.
According to the known GW events, a typical event emits an amount of energy of several solar mass, and the characteristic frequency of the GW is about 100Hz.
Thus for a typical GW event, the number of gravitons is approximately $N~\sim\frac{3\times M_{\bigodot}c^2}{h \nu}=10^{79}$. Now we estate the number of collisions
a graviton. The system of graviton-dark matter is quite different from an ordinary molecular system in thermo equilibrium. The graviton does not walk randomly but walks almost in a strait line, since in every collision it only losses a tiny portion of its momentum. Thus the number of collisions is simplified to the number of dark matter particles in such a cylinder, whose bottom is a disk with radius of the half-wavelength of the graviton, and height is the thickness of the dark matter in consideration. For further calculations, an
 unavoidable problem is that we are almost ignorant about the mass of the dark matter particles. It may be from several TeV to even $10^{-20}$eV. We take 100GeV as an example. Further, we assume the mass of dark matter surrounded is $0.3M_{\odot}$.  As we pointed out in the next section, this is enough to excite the tail wavelet. For more detailed numerical calculations, we assume the thickness of the dark matter is $10^5$km, and distributes homogeneously. With these assumptions, we obtain upper bound the number of collisions of a single graviton is $7\times 10^{53}$, in which we assume the graviton interacts every particle in the travelling cylinder,  and the lower bound is  $10^{17}$, in which we assume the graviton collide only one particle in one step. The total scattering cross section between gravitational radiations and dark matter is therefore, $10^{28}$ to  $10^{64}$m$^2$. As a comparison, the total scattering section of the radiation emitted by the sun in 0.1 second is $10^{16}$m$^2$. The radiation pressure in the interior of the Sun is about $10^{-3}$ of the total pressure. And the momentum per solar photon is $5\times 10^{12}$ times of the momentum per graviton. This is equal to say we need a cross section of of gravitons of $10^{31}$m$^2$ to support a mass as the sun. As we have estimated, the total cross section is about  $10^{28}$ to  $10^{64}$m$^2$.  The accurate value requires the
 details of the interaction between graviton and dark matter particles, which is still unknown. Nevertheless, one sees that it is very possible that the gravitons can push away dark matters around one solar mass for GW150914. Besides, the sunken region of the gravitational potential may concentrate a large number of dark matter particles, as studied in many previous works. Therefore, the total energy transferred from GW to the surrounding dark matter can be considerable.

As a rigorous approach is not yet feasible, in the remainder of the present work, we carry out a phenomenological approach to investigate this interaction between GWs and dark matter.

\textit{II.~Characteristic waveform of the tail wavelets}

The surrounding dark matter may lead to modification of geometry (backreaction). Various modifications have been studied in detail in a fairly sophisticated work \cite{refen}. The result shows that the environmental effects do not spoil the main waves from binary compact objects. In this work we take two typical modifications to
 study possible corrections to the tail wavelet QNMs. The first one is \sch-de Sitter, which describes a black hole with a small amount dark matters (compared to the black hole mass) as a fine approximation,
 \be ds^2=-fdt^2+f^{-1}dr^2+r^2d\Omega^2,~ \en

 where $f=1-2M/r-\Lambda_{\rm eff} r^2/3$. The effective cosmological constant is related to the density of dark matter,
  $ \rho_{DM}=\frac{\Lambda_{\rm eff}}{8\pi}$. We call it SdS case.

   The second one is a short-hair black hole, which describes a non-homogeneous
fluids \cite{short-hair}, where
\be
f=1-2M/r+Q_m^{2k}/r^{2k}.
 \en
 Here $k$ is the equation of state of the parameter of the dark matter. The density and pressure of dark matters are $\rho_m=(2k-1)Q_m^{2k}/(8\pi r^{2k+2})$ and $P_m=k\rho_m$, respectively. The softest matter with $k=-1$ is a cosmological constant, and the most stiff matter with $k=1$ mimics electromagnetic fields.

 Now we estimate the amplitude of the GW around the horizon.
The energy density is proportional to the inverse square of the distance from the source $L$. Thus around the horizon, one has $\frac{\rho_{r_h}}{\rho_{L}}=\frac{L_0^2}{r_h^2}$, where $L_0$ denotes the distance from the Earth to the source of GW150914.
If the energy loss of the GWs is insignificant, the energy density is inversely proportional to the square of the distance from the center, namely, $\frac{A_{r_h}}{A_{L_0}}=\frac{L_0}{r_h}$. For the event GW150914, we have $r_h=1.8\times 10^5$ meters, $L_0=1.2\times 10^{25}$ meters, and $A_{L_0}=10^{-21}$. Here $A_{L_0}$ the amplitude measured on the Earth.
Then one obtains $A_{r_h}=\frac{1}{15}$. Hence it seems to be reasonable to consider that the amplitude of the initial main GWs to be $A_{r_h}\sim 0.1$.

 From the previous studies of tail wavelets (echos), we estate that the amplitude of the wavelet is one tenth of the main wave or smaller. Thus the energy carried by the wavelet is about $0.03M_{\bigodot}$ or smaller. The energy conversion from rest mass energy to gravitational wave energy can be very efficient in the process of splashing down, which is up to $42\%$ \cite{jhartle}. As a conservative estimate, we assume $10\%$ of the total rest mass of the dark matter is transferred to the gravitational energy of the wavelet. Thus the mass of dark matters involved to excite the wavelet is about $0.3M_{\bigodot}$. By using this parameter, we obtain the exact form of backreactions of dark matters to a \sch black hole, and  present the waveform of tail wavelets in the cases of SdS and short-haired black hole, respectively. We find the fine distinction between the cases of a dressed black hole and a bare one, through it is insignificant.

\begin{figure}
\centering
{\includegraphics[width=2in]{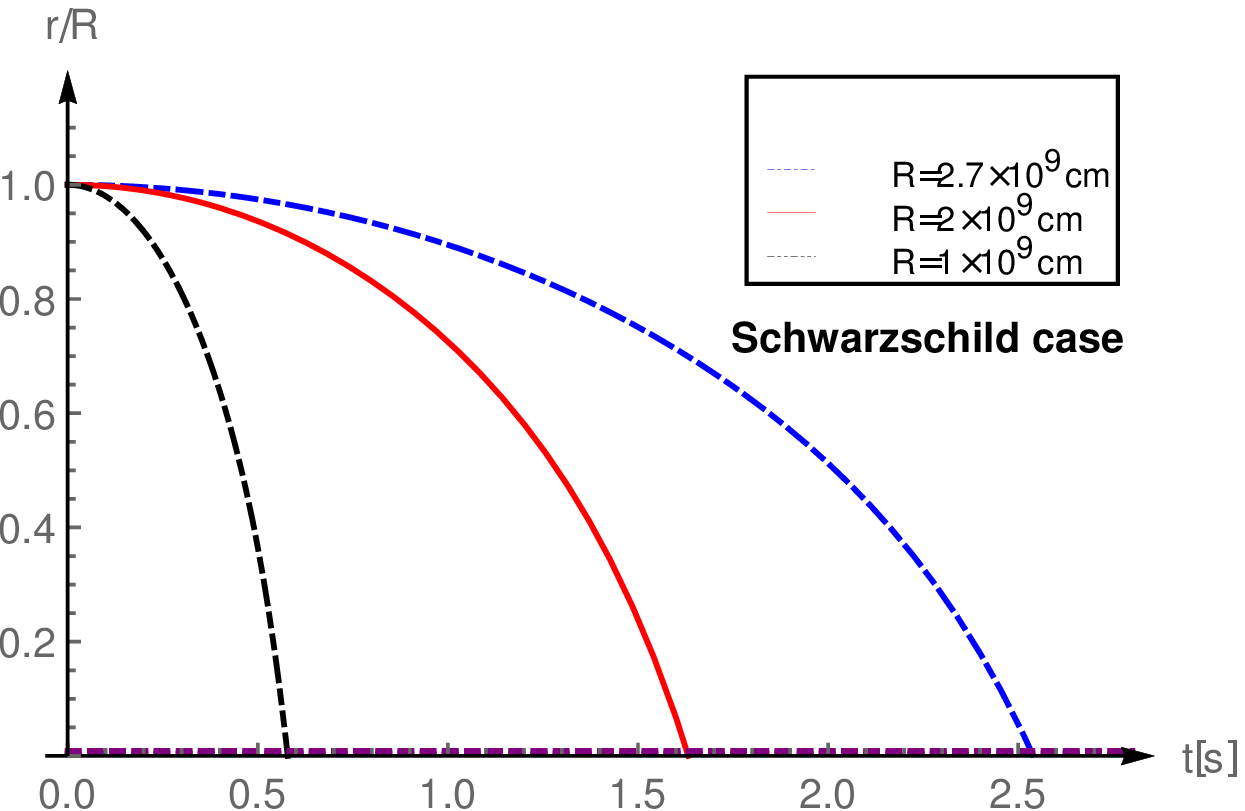}\includegraphics[width=2in]{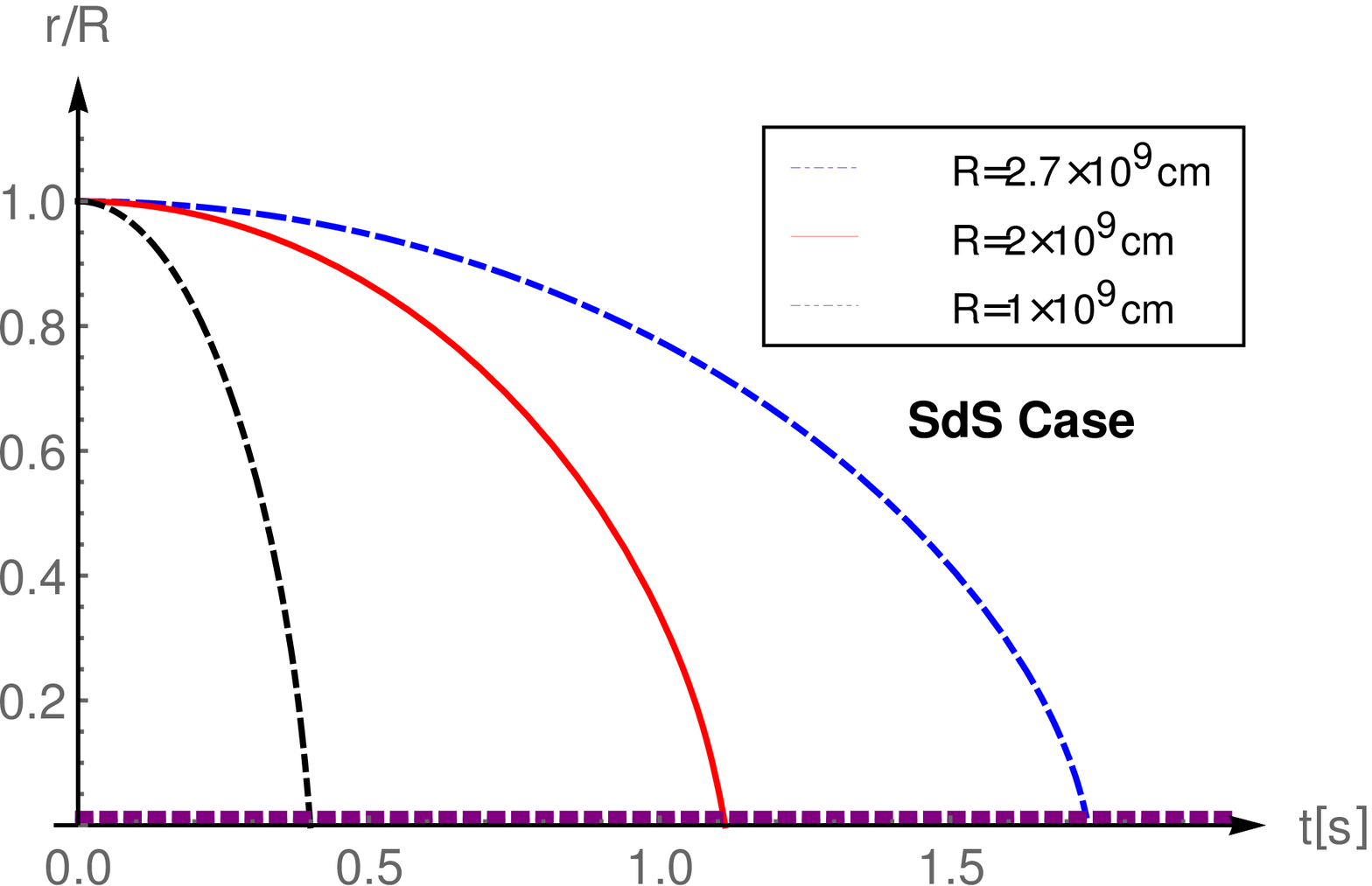}\includegraphics[width=2in]{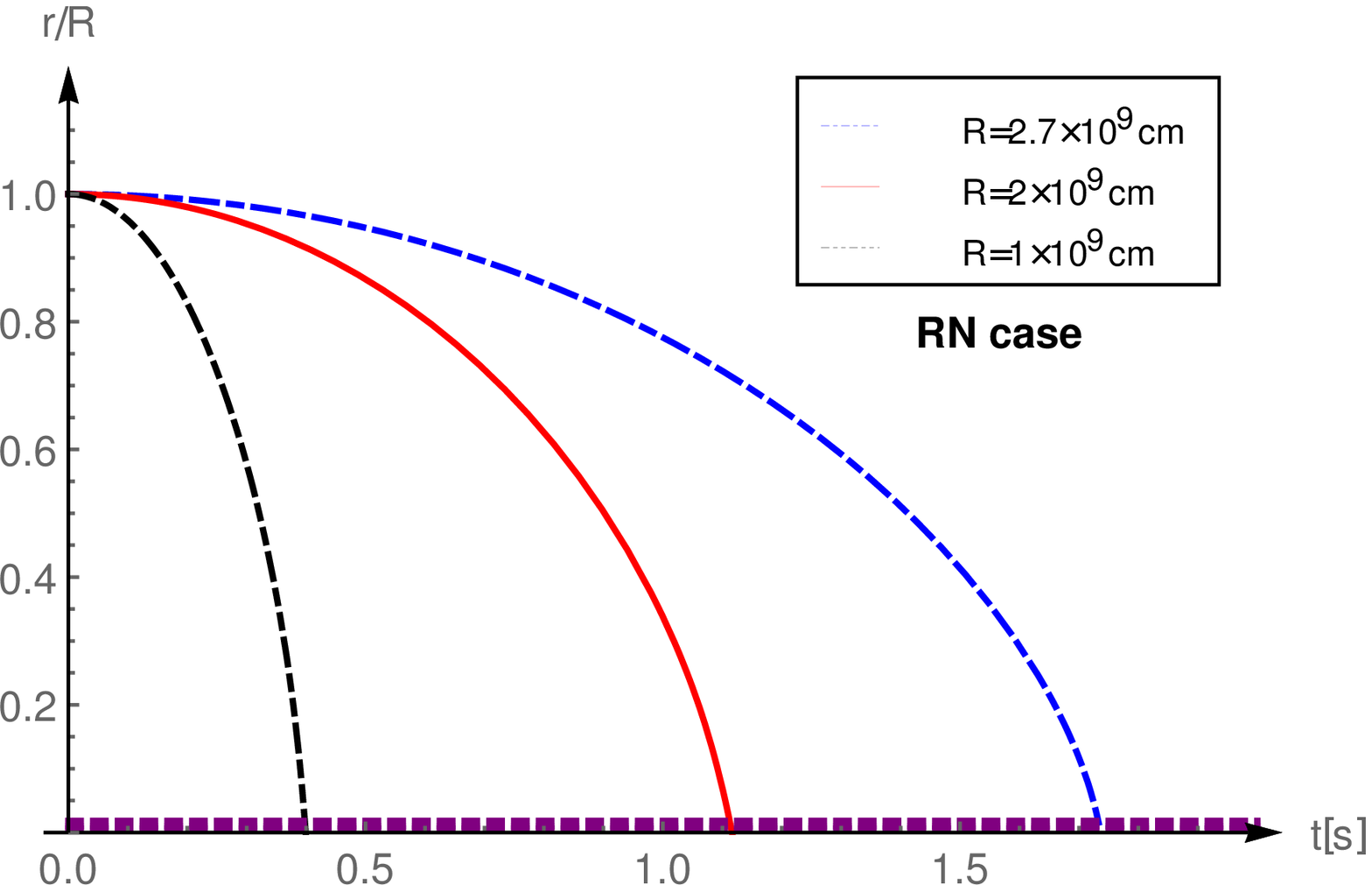}}
\caption{Estimation of epoches of the tail wavelets after the main GWs.
 Left panel: the case without backreaction, i.e., the \sch~. Middle panel: Black hole with DM halos, i.e., the case of SdS. Right panel: Black hole with short-hair.
The location of the event horizon is depicted by the horizontal dashed purple line.}
\label{rRdt}
\end{figure}

  Before showing the waveform of the wavelet, we explore the time scale between the emission of the main GWs and the tail wavelets.
  This can be estimated by simply investigating the equation of motion of a free-falling particle from a given height.
 For simplicity we consider motions along radial geodesics. The results are presented in Fig.\ref{rRdt}.
 There, we consider a particle splashing down from an initial altitude of $R$ towards the horizon $r_h$. One sees that the epoch is advanced a little in the modified geometries. The dark matter halo
  strengthens gravity.

The waveform of the tail wavelet by generated by the infalling of the matter is similar to that of the ordinary quasinormal modes\cite{GQNMs1, GQNMs2}.
Therefore, in our model, the observed tail wavelets are also a manifestation of the black hole quasinormal modes.
  However, there is a subtlety. Since the infalling process is continuous, initial disturbance period carries the information not only on the black hole itself but also about the matter distribution surrounding it.
 It is therefore expected that the waveform in initial disturbance period will be characteristically different from the ordinary quasinormal modes. The Navarro¨CFrenk¨CWhite (NFW) profile is the mostly studied distribution of dark matters around galaxies and clusters \cite{nfw1, nfw2}. As a more realistic study, we consider an NFW distribution of dark matter around the hole. The NFW distribution can be described by,
 \be
 \rho(r)=\frac{\rho_0}{\frac{r}{R_s}\left(1+\frac{r}{R_s}\right)^2}.
 \en
 Setting $x=r/r_h,~K=r_h/R_s$, one obtains,
  \be
 \rho(x)=\frac{\rho_0}{Kx(1+Kx)^2}.
 \en
  Now we assume the dark matter suffuses between $r_h$ and $100r_h$. Thus the total mass of dark matter (including the mass gravitational field) reads \cite{wein},
  \bea
  \nonumber
  M_d=4\pi\int_{r_h}^{100r_h} dr r^2 \rho(r)=4r_h^3\pi\int_{1}^{100}dx x^2 \rho(x)=4\pi r_h^3 \rho_0\int_{1}^{100}dx \frac{x^2}{Kx(1+Kx)^2}\\=4\pi r_h^3 \rho_0 \frac{1}{K^3}(\frac{1}{1+100K}-\frac{1}{1+K}+\ln \frac{1+100K}{1+K}).
  \ena
  We require the total amount of the dark matter equals to the previous homogeneous case, in which the total mass is,
  \be
  M_{dh}=\rho_{0h}4\pi\int_{r_h}^{100r_h} dr r^2=\rho_{0h}\frac{4\pi}{3}10^6 r_h^3.
  \en
  Given a relation between $\rho_0$ and $\rho_{0h}$ one obtains a corresponding $K$. For example, we take $\rho_0=10\rho_{0h}$, and thus we obtain $r_h=0.0254R_s$.

 In this section, we present the numerical calculations of the waveform associated with the gravitational perturbation of the metric owing to the infalling of the matter.

The radial master equation of the gravitational fluctuation around the black hole reads \cite{GQNMs1, GQNMs2},
\be
f(r)\frac{\partial}{\partial r} \left(f(r)\frac{\partial\Phi}{\partial r}\right)-\frac{1}{c^2}\frac{\partial^2\Phi}{\partial t^2}-V(r)\Phi=0.
\en
Here $V$ denotes the potential of the black hole. $V$ is rather complicated. For even and odd modes, the exact form of $V$ see \cite{QNMs1, QNMs2, QNMs3}.
$f(r)$ takes different form in the cases of \sch, SdS, and short-hair black holes, as shown above.

\begin{figure}
\centering
{\includegraphics[width=3in]{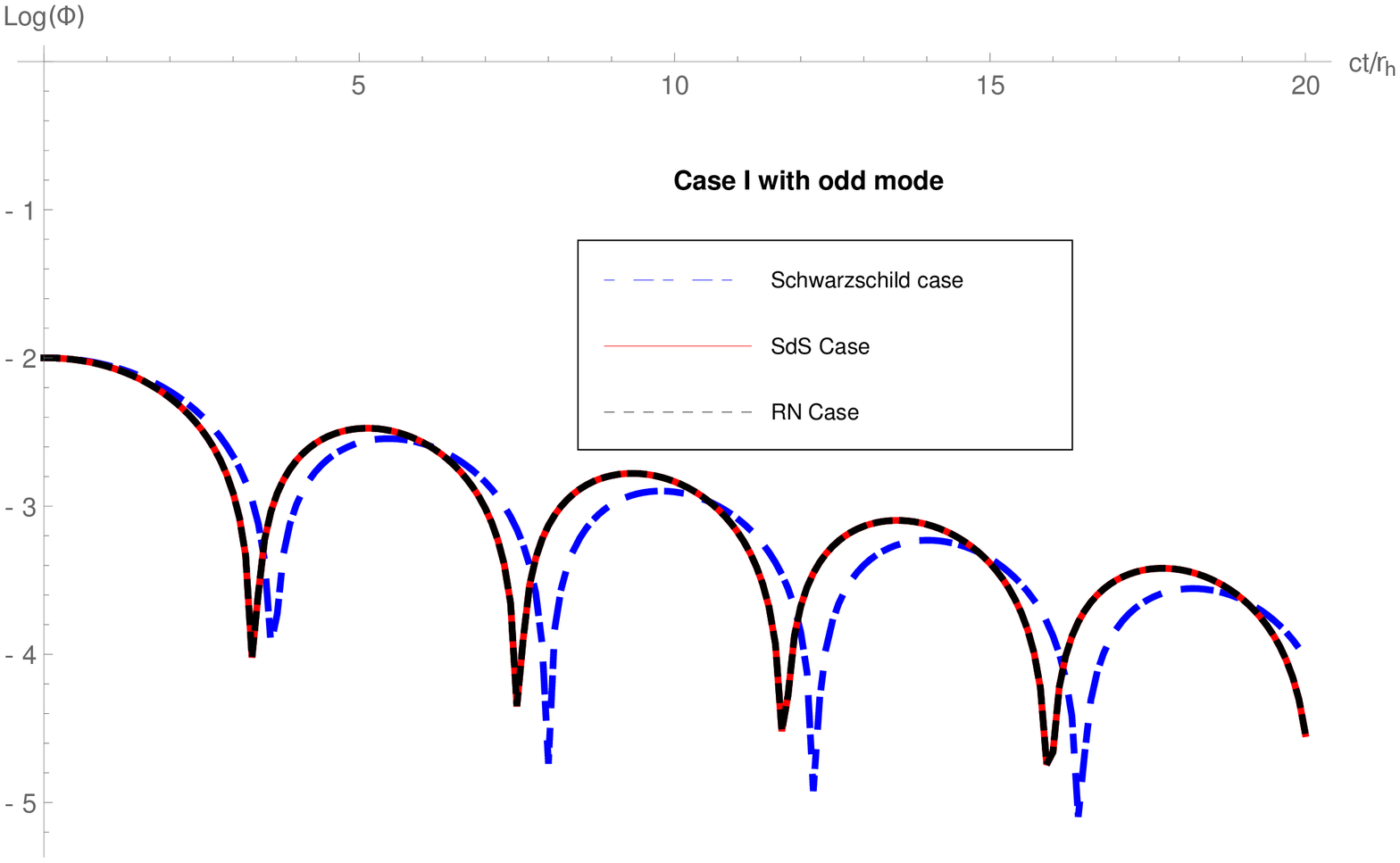}\includegraphics[width=3in]{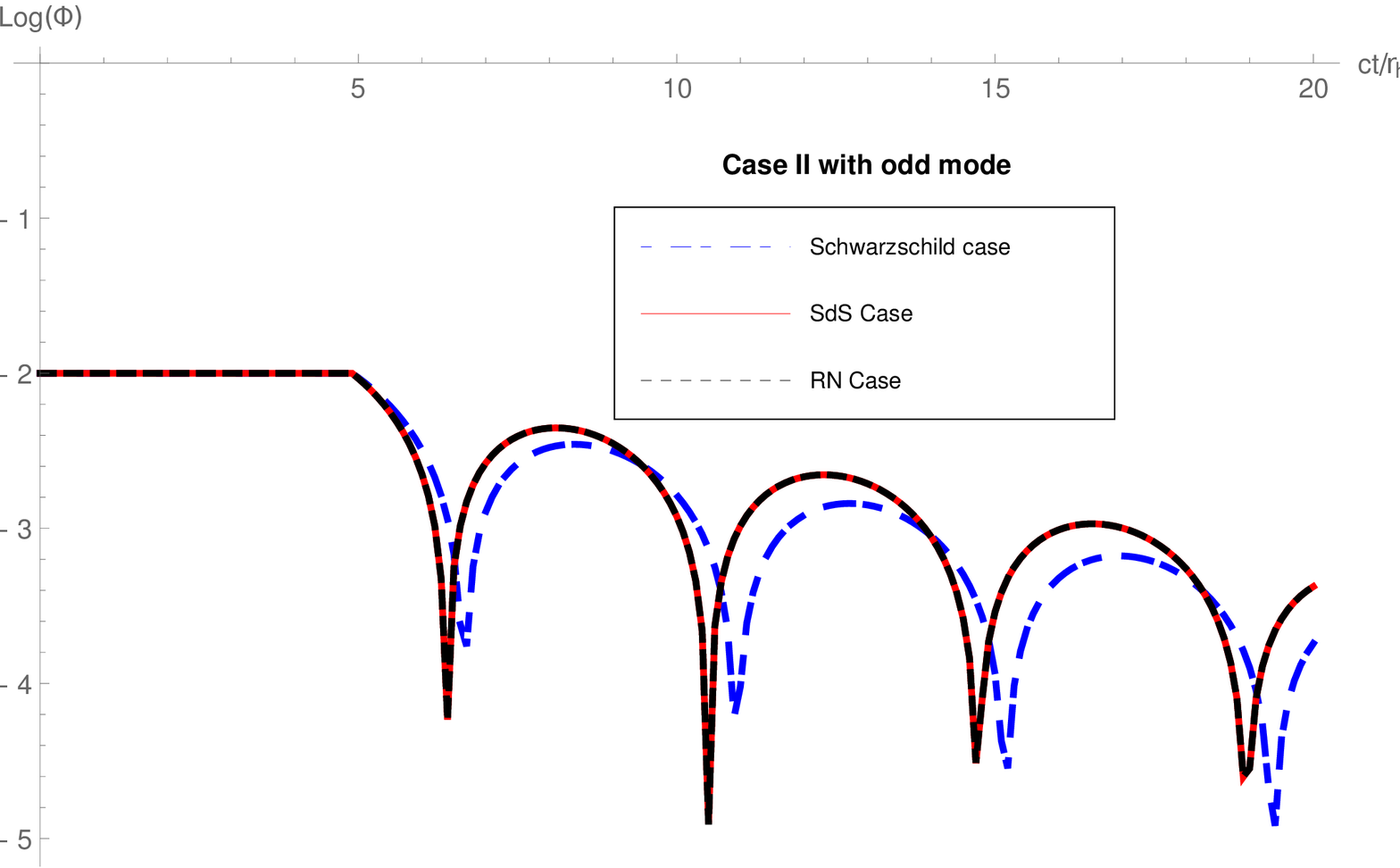}}
\caption{The waveform of odd gravitational fluctuations, including the cases of a bare \sch~black hole and dressed black holes, i.e., SdS and short-haired black hole.}
\label{oddf}
\end{figure}

\begin{figure}
\centering
{\includegraphics[width=3in]{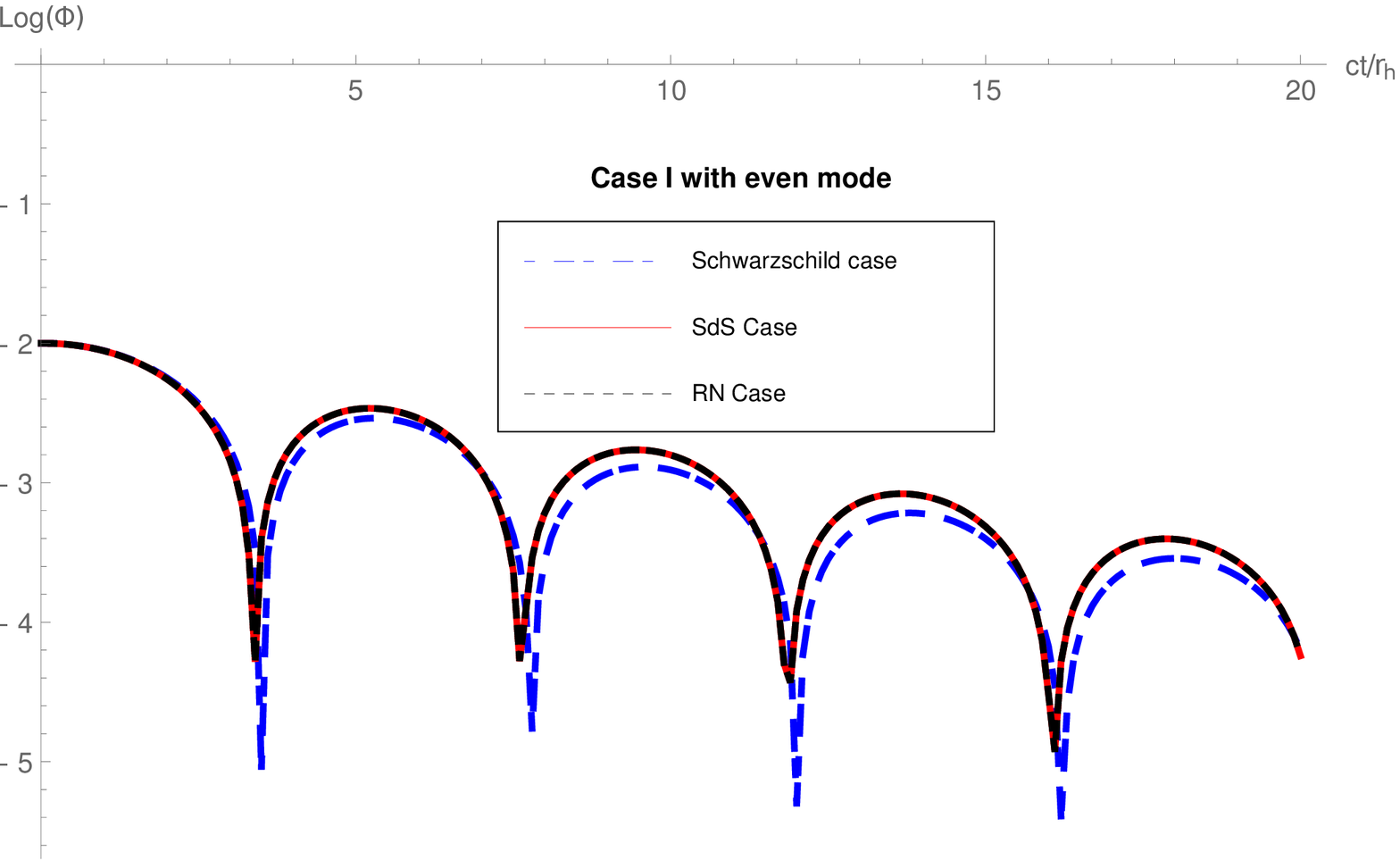}\includegraphics[width=3in]{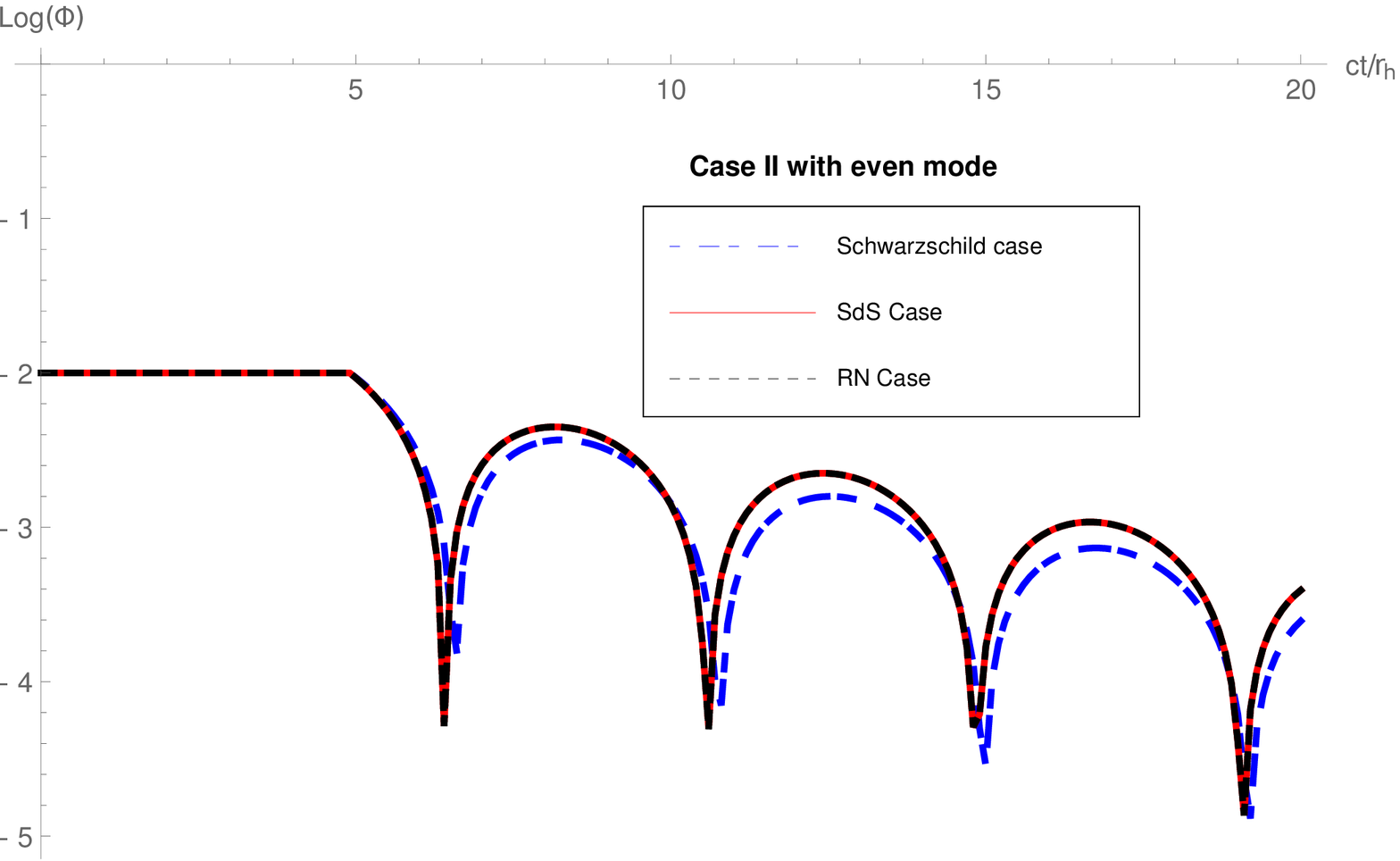}}
\caption{The waveform of even gravitational fluctuations, including the cases of a bare \sch~black hole and dressed black holes, i.e., SdS and short-haired black hole.}
\label{evenf}
\end{figure}

\begin{figure}
\centering
{\includegraphics[width=3in]{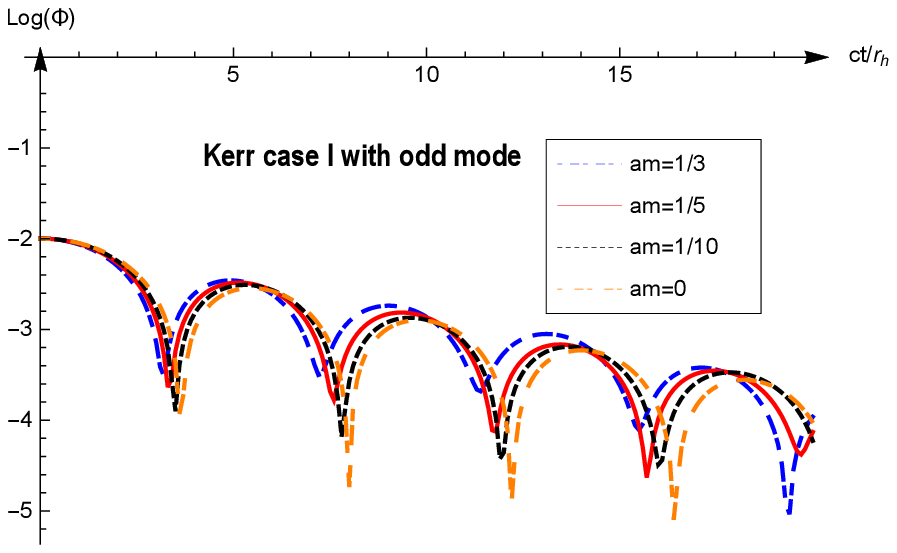}\includegraphics[width=3in]{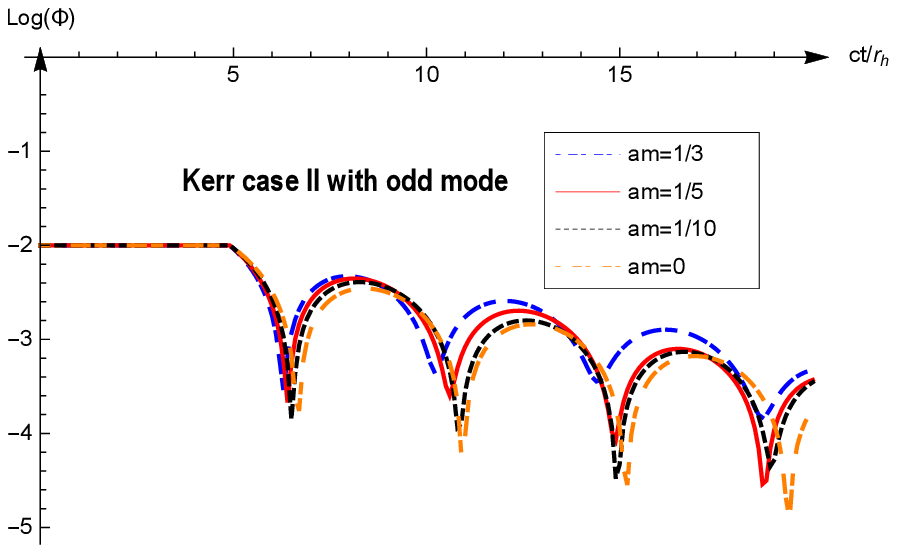}}
\caption{The even modes of the wavelets for rotating black hole. The left panel shows the wavelets corresponding to shell initial
distribution, and the right panel shows the wavelets corresponding to homogeneous distribution. One sees that the even mode
is sensitive to the rotating parameter.}
\label{oddK}
\end{figure}

\begin{figure}
\centering
{\includegraphics[width=3in]{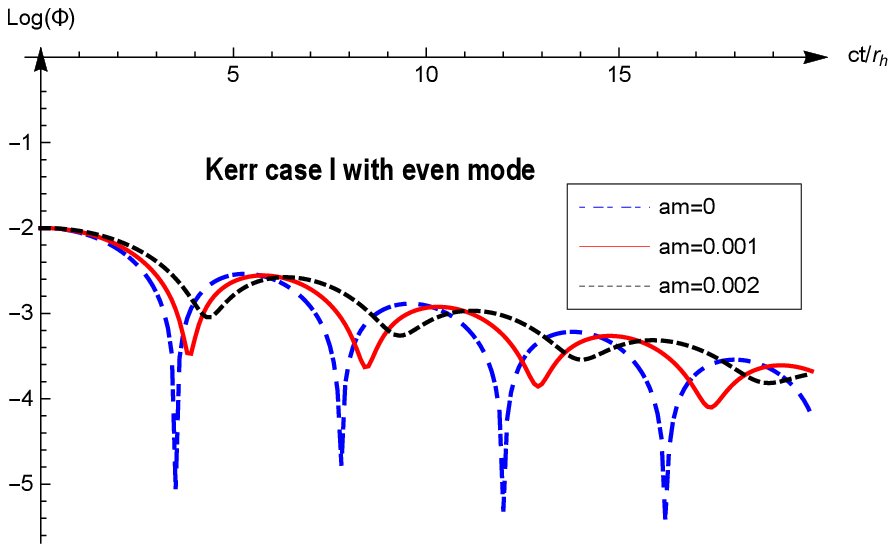}\includegraphics[width=3in]{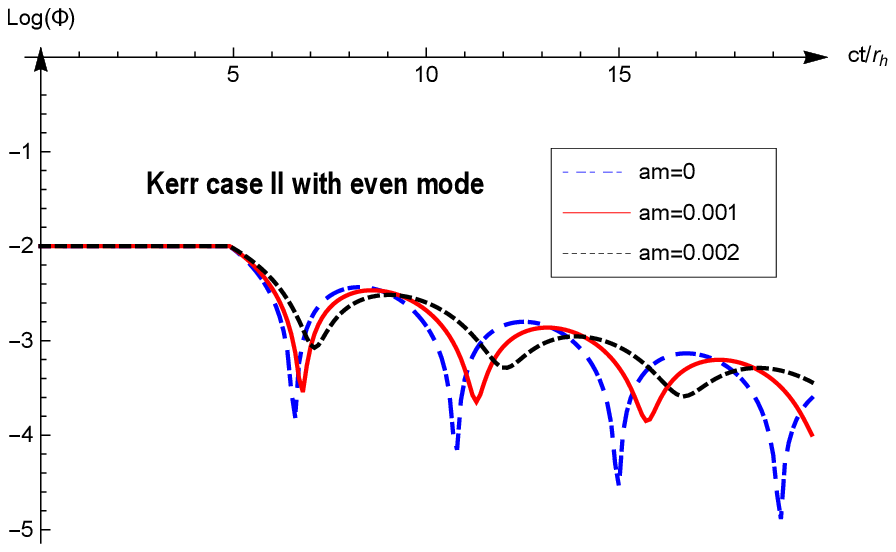}}
\caption{The odd modes of the wavelets for rotating black hole. The left panel shows the wavelets corresponding to shell initial
distribution, and the right panel shows the wavelets corresponding to homogeneous distribution. One sees that the odd mode
is not sensitive to the rotating parameter.}
\label{evenK}
\end{figure}

Based on the analysis in the previous sections, we take the following initial conditions.
For convenience, we choose $\eta=ct/r_h$ and consider the following different cases:

Case I, $\Phi=0.01$ as $\eta \in (0,5)$.

Case II, $\Phi=0.01$ as $\eta=0$

The first case denotes dark matter initially in a homogeneous distribution. As a comparison, we plot the second case, which displays a pulse excitation which is the ordinary QNM of black hole.
Figs.\ref{oddf} and \ref{evenf} demonstrate the waveforms of the odd and even fluctuations for all the cases of geometries, including the \sch, the geometry with backreaction of dark matter halos (SdS), and the geometry of a short-hair black hole. According to the present estimations, the magnitude of the tail wavelets is at most $\sim$10\% of the main GWs \cite{renjing,duchen}.
Thus we take the initial perturbation to be $\Phi=0.01$ for our calculations.
One observes that the waveforms generated by some distribution of dark matters are different from those trigger by a pulse, the latter corresponds to the ordinary quasinormal mode which only carries the information of the black hole itself. Further, one sees that the effects of modifications of geometry is not significant compared to the case of a \sch~black hole. Especially, the two modifications present almost the same wavelet, which is slightly different from a ``bare" \sch~one.

 We expect forthcoming data with better resolution for the waveform, especially those from the third generation GW detectors like the Einstein Telescope \cite{et1, et2} and Cosmic Explorer \cite{ce}, may shed light on the feasibility of the present model.

Generally, the astrophysical black hole is Kerr black hole. We study the Schwarzschild black hole is as an approximation
of the realistic case. Now we present a preliminary investigation of rotating black holes. We plot the
corresponding wavelets for Kerr holes in fig 4 and fig 5.



{\textit{III.~Concluding remarks}}

Data analysis implies that there might be tail wavelets occurring after the main GWs from the merger of binary
systems. Such tail wavelets exist not only for the GW events of binary black holes but also for those of binary neutron
stars. Therefore, it is speculated that the physical mechanism behind the phenomenon might not be related to the
quantum structure of the black hole horizon but is rather associated with certain generic properties of the collapsing
binary system. The main characteristic of our approach is that the explanation is given within the framework of
Einstein¡¯s general relativity, rather than originated from a modified theory of gravity or quantum effect.

Following this line of thought, in this work, we present a more natural and straightforward scenario for the generation of the tail wavelets. In our model, the cause of the phenomenon is attributed to the matter or dark matter surrounding
the binary system. In particular, the dark matter distribution around a compact object has been studied in different
scales for many years. We carefully investigate the pressure of the GWs and demonstrate that the GWs can perturb
and even push away the matter distributed around the compact star. After the main wave traverses through, the
matter falls back towards the black hole horizon, which in turn excites the tail wavelets. Based on our analysis,
we assume four different initial conditions for the related gravitational perturbations and evaluate the corresponding
waveforms. Furthermore, we make a preliminary exploration of the wavelet for Kerr black holes.

In this paper we concentrate on the physics after inspirals. We explore the effects of energy emitting in the merge process, and especially the splashing down of dark matter to the black hole.  It is not closely related to the physics before merging. According to the previous studies of intermediate-mass-ratio inspirals (IMRI) without dark matter \cite{hancaohu}, and especially an upcoming study \cite{hanchen} there is almost no wavelet in IMRI. However, we show that dark matter in such a system can produce significant tail wavelet in the merger of binary compact objects. The effect of dark matter in the inspiraling stage of IMRI system has been investigated in \cite{hanyue}, which demonstrates that dark matter significantly enhances the merging process. We show that dark matter has significant effect after merging process.

We look forward to testing the proposed model against the forthcoming data from the GW detectors of the third
generation. Further studies concerning specific dark matter distributions are in progress. Furthermore, in frame of this
model one can investigate the distribution of dark matter surrounding the black holes through analysing the waveform
of the tail wavelet from the forthcoming data with better resolution, especially those from the third generation GW
detectors.

\section*{Acknowledgments}
Our hearts felt thanks to Professor Anzhong Wang, Zhoujian Cao and Wenbiao Han for helpful discussions.
This work is supported in part by the National Natural Science Foundation of China (NSFC) under grant Nos.11805166 and 11575083, as well as Shandong Province Natural Science Foundation under grant No. ZR201709220395.
We also gratefully acknowledge the financial support from Brazilian funding agencies Funda\c{c}\~ao de Amparo \`a Pesquisa do Estado de S\~ao Paulo (FAPESP),
Conselho Nacional de Desenvolvimento Cient\'{\i}fico e Tecnol\'ogico (CNPq), Coordena\c{c}\~ao de Aperfei\c{c}oamento de Pessoal de N\'ivel Superior (CAPES).
X.F. is supported by the National Natural
Science Foundation of China under grants No. 11673008 and
11633001, the Strategic Priority Program of the Chinese
Academy of Sciences (grant No. XDB 23040100) and Newton
International Fellowship Alumni Follow-on Funding.


\begin{thebibliography}{99}

 \bibitem{MTW}
 C Misner, K Thorne, J Wheeler, $Gravitation$, San Francisco, USA, W. H. Freeman(1973).
  \bibitem{fuzzb}
 O. Lunin and S. D. Mathur, Phys. Rev. Lett. 88, 211303
(2002), arXiv:hep-th/0202072 [hep-th].
\bibitem{firewall1}
A. Almheiri, D. Marolf, J. Polchinski, and J. Sully, JHEP
02, 062 (2013), arXiv:1207.3123 [hep-th].
 \bibitem{firewall2}
 H.~Zhang,
  Sci.\ Rep.\  {\bf 7} (2017) no.1,  4000
  doi:10.1038/s41598-017-03854-y
  [arXiv:1706.08850 [gr-qc]].

 \bibitem{GWs}
    B.~P.~Abbott {\it et al.} [LIGO Scientific and Virgo Collaborations],
  Phys.\ Rev.\ Lett.\  {\bf 116}, no. 6, 061102 (2016)
  doi:10.1103/PhysRevLett.116.061102
  [arXiv:1602.03837 [gr-qc]]; for all known GW events, see https://en.wikipedia.org/wiki/List\_of\_gravitational\_wave\_observations

\bibitem{echo1}
  J.~Abedi, H.~Dykaar and N.~Afshordi,
  Phys.\ Rev.\ D {\bf 96}, no. 8, 082004 (2017)
  doi:10.1103/PhysRevD.96.082004
  [arXiv:1612.00266 [gr-qc]].
  \bibitem{echo2}
  V.~Cardoso and P.~Pani,
  arXiv:1707.03021 [gr-qc].
  \bibitem{echo31}
  V. Cardoso, E. Franzin, and P. Pani,
   Phys. Rev. Lett. 116 no. 17,
(2016) 171101, arXiv:1602.07309 [gr-qc].
 \bibitem{echo32}
 L.~Barack {\it et al.}, arXiv:1806.05195 [gr-qc].

  \bibitem{noecho1}
  W. Julian  {\it et al.}     Physical Review D, 97 (124037) arXiv:1712.09966, (2018).
  
  \bibitem{noecho2}
A. B. Nielsen,   {\it et al.} Physical Review D, 99 (104012) arXiv:1811.04904, (2019).

 \bibitem{renjing}
  R.~S.~Conklin, B.~Holdom and J.~Ren,
  Phys.\ Rev.\ D {\bf 98}, no. 4, 044021 (2018)
  doi:10.1103/PhysRevD.98.044021
  [arXiv:1712.06517 [gr-qc]].
   \bibitem{duchen}
  S.~M.~Du and Y.~Chen,
  Phys.\ Rev.\ Lett.\  {\bf 121}, no. 5, 051105 (2018)
  doi:10.1103/PhysRevLett.121.051105
  [arXiv:1803.10947 [gr-qc]].


  \bibitem{NStarwave}
  J.~Abedi and N.~Afshordi,
  [arXiv:1803.10454 [gr-qc]].

  \bibitem{NStarwave2}
  P.~Pani and V.~Ferrari,
  Class.\ Quant.\ Grav.\  {\bf 35}, no. 15, 15LT01 (2018)
  doi:10.1088/1361-6382/aacb8f
  [arXiv:1804.01444 [gr-qc]].


 \bibitem{piwork1}
  C.~S.~Frenk, S.~D.~M.~White, M.~Davis and G.~Efstathiou,
  Astrophys.\ J.\  {\bf 327}, 507 (1988).
  
   \bibitem{piwork2}  
   J.~Dubinski and R.~G.~Carlberg,
  Astrophys.\ J.\  {\bf 378}, 496 (1991).


 \bibitem{nfw1}
J.~F.~Navarro, C.~S.~Frenk and S.~D.~M.~White,
  Astrophys.\ J.\  {\bf 490}, 493 (1997)
  doi:10.1086/304888
  [astro-ph/9611107].
  
  \bibitem{nfw2}
  J.~F.~Navarro, C.~S.~Frenk and S.~D.~M.~White,
  Astrophys.\ J.\  {\bf 462}, 563 (1996)
  doi:10.1086/177173
  [astro-ph/9508025].
  \bibitem{sbh1}
     D.~Merritt,
  Phys.\ Rev.\ Lett.\  {\bf 92}, 201304 (2004)
  doi:10.1103/PhysRevLett.92.201304
  [astro-ph/0311594].
  \bibitem{sbh2}
   C. M. Booth and J. Schaye, Mon.Not.Roy.Astron.Soc. 405: L1(2010).
  \bibitem{disneu1}
   J.~Ellis, A.~Hektor, G.~H\"utsi, K.~Kannike, L.~Marzola, M.~Raidal and V.~Vaskonen,
  Phys.\ Lett.\ B {\bf 781} (2018) 607
  doi:10.1016/j.physletb.2018.04.048
  [arXiv:1710.05540 [astro-ph.CO]].
  \bibitem{disneu2}
   S.~C.~Leung, M.~C.~Chu and L.~M.~Lin,
  Phys.\ Rev.\ D {\bf 84}, 107301 (2011)
  doi:10.1103/PhysRevD.84.107301
  [arXiv:1111.1787 [astro-ph.CO]].
  
  \bibitem{disneu3}
    A.~Li, F.~Huang and R.~X.~Xu,
  Astropart.\ Phys.\  {\bf 37}, 70 (2012)
  doi:10.1016/j.astropartphys.2012.07.006
  [arXiv:1208.3722 [astro-ph.SR]].
  \bibitem{hol1}
  B.~R.~Holstein,
  Am.\ J.\ Phys.\  {\bf 74}, 1002 (2006)
  doi:10.1119/1.2338547
  [gr-qc/0607045].
  \bibitem{hol2}
   T.~Rothman and S.~Boughn,
  Found.\ Phys.\  {\bf 36}, 1801 (2006)
  doi:10.1007/s10701-006-9081-9
  [gr-qc/0601043].
  \bibitem{dyson}
  F.~Dyson,
  Int.\ J.\ Mod.\ Phys.\ A {\bf 28}, 1330041 (2013).
  doi:10.1142/S0217751X1330041X




\bibitem{refen}
E.~Barausse, V.~Cardoso and P.~Pani,
  Phys.\ Rev.\ D {\bf 89}, no. 10, 104059 (2014)
  doi:10.1103/PhysRevD.89.104059
  [arXiv:1404.7149 [gr-qc]].
  \bibitem{short-hair}
  J. D. Brown and V. Husain, Int. J. Mod. Phys. D 06, 563
(1997).

\bibitem{jhartle}
J. Hartle, $Gravity:~ An ~Introduction ~to~ Einstein's~ General~ Relativity$, San Francisco: Addison Wesley Press(2003), page 321.

\bibitem{wein}
  S.~Weinberg,
  $Gravitation~ and ~Cosmology :~ Principles~ and~ Applications~ of~ the~ General~ Theory~ of~ Relativity$, New York: John Wiley and Sons (1972).



\bibitem{GQNMs1}
T. Regge and J. A. Wheeler,  Physical Review, {\bf 108} 1063-1069, (1957).
\bibitem{GQNMs2}
F. J. Zerilli, Physical Review Letters, {\bf 24} 737-738, (1970).
\bibitem{QNMs1}
R Konoplya and A Zhidenko, Rev. Mod. Phys. 83 793 (2011).
\bibitem{QNMs2}
R A Konoplya, Phys. Rev. D 68, 124017 (2003).
\bibitem{QNMs3}
K Lin, W-L Qian and A. B. Pavan, Phys. Rev. D94 064050 (2016).

\bibitem{et1}
S Hild  et al 2011 Class. Quantum Grav. 28 094013.
\bibitem{et2}
M Abernathy et al and ET Science Team 2010 Technical Report No. ET-0106C-10.
\bibitem{ce}
B Abbott et al., 2017, Class. Quantum Gravity, 34, 044001.

\bibitem{hancaohu}
  W.~B.~Han, Z.~Cao and Y.~M.~Hu,
  Class.\ Quant.\ Grav.\  {\bf 34} (2017) no.22,  225010
  doi:10.1088/1361-6382/aa891b
  [arXiv:1710.00147 [gr-qc]].

  \bibitem{hanchen}
  Wen-Biao Han, Yanbei Chen et al, to appear.

  \bibitem{hanyue}
  X.~J.~Yue, W.~B.~Han and X.~Chen,
  Astrophys.\ J.\  {\bf 874}, no. 1, 34 (2019)
  doi:10.3847/1538-4357/ab06f6
  [arXiv:1802.03739 [gr-qc]].


\end{thebibliography}
\end{document}